\documentclass[12pt]{article}

\usepackage[english]{babel}
\usepackage[utf8x]{inputenc}
\usepackage[T1]{fontenc}

\usepackage[a4paper,top=2.5cm,bottom=2cm,left=2.5cm,right=2.5cm,marginparwidth=1.75cm]{geometry}

\usepackage{amsmath}
\usepackage{graphicx}
\usepackage[colorinlistoftodos]{todonotes}
\usepackage[colorlinks=true, allcolors=blue]{hyperref}

\title{CubeSats for Astronomy and Astrophysics\\
\small{submitted to the Astro 2020 Decadal Survey Call for APC whitepapers}}
\date{10 July 2019}
\begin{document}
\maketitle
\begin{center}
    
Ewan S. Douglas$^1$,
Kerri L. Cahoy$^{3,4}$,
Mary Knapp$^2$,
Rachel E. Morgan$^3$.
\\
$^1$Department of Astronomy and Steward Observatory, University of Arizona;\\
$^2$Haystack Observatory, MIT;\\
$^3$Department of Aeronautics and Astronautics, MIT;\\
$^4$Department of Earth, Atmospheric, and Planetary Science, MIT.\\
\end{center}
Co-Signers:\\
Supriya Chakrabarti - UMASS Lowell. \\
Timothy Cook - UMASS Lowell.\\
Evgenya Shkolnik - Arizona State University.





\section{Introduction}
CubeSats have the potential to expand astrophysical discovery space, complementing ground-based electromagnetic and gravitational-wave observatories. The CubeSat design specifications\footnote{http://www.cubesat.org} help streamline delivery of instrument payloads to space. 
CubeSat planners have more options for tailoring orbits to fit observational needs and may have more flexibility in rapidly rescheduling observations to respond to transients. With over 1000 CubeSats launched\footnote{As of June 2019, \url{https://sites.google.com/a/slu.edu/swartwout/home/cubesat-database}}, there has been a corresponding increase in the availability and performance of commercial-off-the-shelf (COTS) components compatible with the CubeSat standards, from solar panels and power systems to reaction wheels for three axis stabilization and precision attitude control. Commercially available components can reduce the cost of CubeSat missions, allowing more resources to be directed toward scientific instrument payload development and technology demonstrations. 

We will discuss the value already provided by current missions, describe several other already-selected  astrophysics observation and technology demonstration CubeSat flight missions, discuss other relevant astrophysics applications for CubeSats such as gravitational wave event follow-up, and assess what additional instrumentation and technology development areas, such as propulsion and high-data rate communications, are still needed in order to maximize benefit from a CubeSat astrophysics platform.
For additional details on astrophysical potential of CubeSats see Shkolnik \cite{2018NatAs...2..374S}.
\textbf{We endorse the findings of the the 2016 report ``Achieving Science with CubeSats'' \cite{national_academies_of_sciences_achieving_2016} and expand on them: suggesting the consideration of several details including increased mission cadence, standard spacecraft buses, and streamlining of licensing.}

\section{Recent Progress in CubeSats for Astronomy}

Despite the severe restrictions that CubeSats have in cost, size, weight, and power (SWaP) which also limit aperture size without the development and use of more complex deployables, CubeSat constellations or swarms can still enable improved spectral, temporal, and spatial coverage of astrophysical targets. CubeSats can also be used for technology demonstrations; while such demonstrations may not be able to make the desired scientific observations on a CubeSat platform, they can reduce risk and increase the technology readiness level of key elements of future scientific instruments, such as detectors, actuators, optical sub-assemblies, and drive electronics.
Examples of astrophysics nanosatellite flight missions that have already launched include the Cosmic X-Ray Background Nanosatellite (CXBN) mission \cite{2012SPIE.8507E..19S}, the Bright Target Explorer (BRITE) (Baade, D. et al. Short-term variability and mass loss in Be stars - I. BRITE satellite photometry of $\eta$ and $\mu$ Centauri. A\&A 588, A56 (2016)) and the Arcsecond Space Telescope Enabling Research in Astrophysics (ASTERIA) \cite{smith_-orbit_2018}  CXBN-2 was launched in 2012 to observe the diffuse X-ray background and help understand the underlying physics of the early universe. The BRITE constellation of nanosatellites has demonstrated millimagnitude photometry of Be stars.

\subsection{ASTERIA}
 The Arcsecond Space Telescope Enabling Research in Astrophysics (ASTERIA), delivered and launched in 2017 by NASA JPL in collaboration with MIT, is designed to demonstrate key technologies needed for high precision (sub-mmag) photometric precision in a small package: high precision line-of-sight pointing control and thermal stability.  ASTERIA demonstrated 0.5 arcsec RMS line-of-sight pointing control and $\pm$10 mK thermal stability \cite{smith_-orbit_2018}. 
 
 ASTERIA's payload instrument is composed of a broadband visible light CMOS detector, refractive optics (f/1.6), and a short baffle to reduce stray light.  The payload takes up approximately 2.5U of the 6U spacecraft.  ASTERIA collected photometric timeseries data both during ASTERIA's three-month prime mission and subsequent mission extensions. One of the observational targets was 55 Cnc, a nearby Sun-like star with several known exoplanets.  The innermost planet in the system, 55 Cnc e, is a transiting 2R$_{Earth}$ super Earth with an 18 hour orbital period and transit depth of $\sim$400 parts per million (ppm).  ASTERIA detected the transit of 55 Cnc e at 3$\sigma$.  This is the first CubeSat measurement of an exoplanet transit.

\section{Upcoming CubeSat Missions}
\subsection{Science}
\subsubsection{SPARCS}
The Star-Planet Activity Research CubeSat (SPARCS) is a NASA-funded astrophysics mission, devoted to the study of the ultraviolet (UV) monitoring of low-mass stars. Given their large abundance and small size, low-mass stars are prime targets for the search for and characterization of habitable zone (HZ) terrestrial exoplanets. Understanding the environmental conditions of a HZ planet requires the assessment of the central star’s UV emission and variability. However, not enough is known about these stars’ high-energy variability, flaring, and quiescent emission, which power photoevaporation and photochemical reactions in the atmospheres of planets. Over its initial 1-year mission, SPARCS will stare at ≈12 M stars of various ages (10 Myr to 10 Gyr) in order to measure short-tefm (over minutes) and long-term (over weeks)  variability simultaneously in the near-UV (NUV, $\lambda_c$= 280 nm) and far-UV (FUV, $\lambda_c$  = 162 nm). The SPARCS payload consists of a 9-cm reflector telescope coupled with two UV optimized 2D-doped CCDs which have quantum efficiencies 5-7 times that of past detectors. The payload will be integrated within a 6U CubeSat, which is slated to be launched as part of a ride-share program into a Sun-synchronous terminator orbit, allowing for long observing stares for all targets. The spacecraft will be ready for launch by the end of 2021. (See Shkolnik et al. 2018, Scowen et al. 2018, Ardila et al. 2018)

\subsection{Other missions}
The NASA funded HaloSat mission  to search for missing baryons was successfully deployed, and has started to release first results \cite{2019AAS...23346206J}. 
The Colorado Ultraviolet Transit Experiment (CUTE) is a 4- year NASA funded near-ultravioplet mission to perform transit spectroscopy of exoplanet atmospheres\footnote{http://lasp.colorado.edu/home/cute/}.

\subsection{Technology}
\subsubsection{DeMi}

The Deformable Mirror Demonstration Mission (DeMi) is a 6U CubeSat mission that will demonstrate a Microelectromechanical Systems (MEMS) Deformable Mirror (DM) in space \cite{morgan_mems_2019}. MEMS DMs are a promising technology to enable the extreme wavefront control required for direct imaging of exoplanets with future space telescopes. The optical payload is essentially a miniature space telescope with an adaptive optics instrument. The payload can observe either external stellar targets or an internal laser calibration source, and can measure the optical wavefront in the system with a Shack Hartmann wavefront sensor and an image plane camera. DeMi is an example of a CubeSat with the purpose of raising technology readiness level (TRL).


\section{Prospects for the future}

 In the era of time domain astronomy, rapid, multi-wavelength follow-up of transient targets in the ultraviolet (UV) and X-Ray is of particular interest.
 For example, recent  UV  observation of electromagnetic emission coincident with gravitational-wave detection of a binary neutron star coalescence was recorded by the 30 cm SWIFT UV Optical Telescope with 120 second observations\cite{evans_swift_2017-1}. 
 A constellation of independent CubeSats with somewhat longer exposure times could provide simultaneous  follow-up to multiple such events.

\section{Recommendations}

Missions like ASTERIA  have demonstrated CubeSat's potential to complement traditional suborbital balloon and sounding rocket programs by demonstrating longer mission duration, longer exposure times, and higher altitudes at the expense of re-usability and aperture.

The 2016 National Academies report ``Achieving Science with CubeSats''\\ \cite{national_academies_of_sciences_achieving_2016}  had three conclusions related to astrophysics which we endorse: 
\begin{quote}
    
1. Observations of variable sources including variable stars and transiting planets—a CubeSat can stare for long time periods at targets of interest, for example; 2. Interferometry—CubeSats can form swarms and arrays that create new opportunities for multi-aperture observations; 3. Technology de-risking—CubeSats can be platforms for new technology development and testing of sensors and and system methodologies that will enable larger missions.''
\end{quote}

Additionally, we make the recommendations below.

\subsection{Study of a NASA administered CubeSat bus}
Spacecraft bus innovation is a key strength of the CubeSat platform; however, making CubeSats a routine platform for astrophysics is limited by logistical barriers such as FCC licensing of each radio and the systems engineering required to adapt to different electrical and mechanical CubeSat bus configurations.  The suborbital community benefits significantly from the standardization of communication interfaces such as WFF 93 and a detailed, publicly available handbook for sounding rockets\footnote{\url{https://sites.wff.nasa.gov/code810/files/SRHB.pdf}}. 
Similarly, most astrophysics CubeSats are expected to require precision attitude control and high data rates.
\textbf{We propose a study the viability of a standard bus or buses (e.g. 3U and 6U) which can be proposed for astrophysics missions.}

One way of implementing this would be a few dedicated "families" of CubeSats that are consistently developed by NASA as preferred buses like the NASA Rapid Spacecraft Development Office catalog for larger missions\footnote{https://rsdo.gsfc.nasa.gov/catalog.html}, to minimize redundant effort for individual missions reinventing the wheel and redefining common interface documents. Supporting open source software repositories, such as the core Flight System (cFS) developed by NASA Goddard, that make flight software more modular and reusable will also increase the accessibility of CubeSats to the astrophysics community.

\subsection{Licensing}
Licensing of CubeSat radios and the scheduling of the NASA Wallops ground station facility have become  significant logistical challenges for CubeSat missions which might be mitigated by standardized high data-rate radios and expansion of federally supported ground stations. These groundstations could be university run to  maximize student training, geographic diversity, and redundancy -- on the model of the many existent UHF groundstations. UMASS Lowell, MIT,  and U. of Arizona are all working to fill this gap with X-band assets. We also propose study of a licensing process which is handled by NASA/NSF rather than individual CubeSat PIs to reduce administrative challenges. 

\subsection{Increase Investment in Enabling Technologies}
Propulsion, high-bandwidth/long distance communications, and precision attitude determination and control (ADCS) are all required to fully realize CubeSats’ science potential. 
While recent advances have brought CubeSat pointing to arcsecond scales, investment in higher precision and lower SWAP ADCS systems will allow larger astronomy payloads with longer exposure times.

Propulsion is an extremely important investment for astronomy missions. NASA should support development of chemical and electric propulsion systems for CubeSats.
Propulsion combined with precision ADCS will enable powerful constellations of CubeSats.

Future astrophysics CubeSats will benefit from operation over longer mission lifetimes and in orbits with less benign radiation environments than low Earth orbit, which will require either development of radiation tolerant or hardened CubeSat components, comprehensive testing and qualification of COTS CubeSat components for operation in harsher environments, or both.

\subsection{Increase Individual Mission Cadence}
CubeSats provide significant workforce training opportunities and CubeSat constellations have significant science potential.
We echo the \cite{national_academies_of_sciences_achieving_2016} recommendation:
\begin{quote}
    NASA should use CubeSat-enabled science missions as hands-on training opportunities to develop principal investigator leadership, scientific, engineering, and project management skills among both students and early career professionals. NASA should accept the risk that is associated with this approach..
\end{quote}
We encourage a strong astrophysics CubeSat program to test individual, high-risk payloads with do-no-harm mission assurance requirements in addition to large constellation missions managed at higher mission assurance levels. 

Some missions, particularly unique high-risk missions, as well as those closely related to a ground-based observatory such as for LSST followup, may be well suited to National Science Foundation funding. 

\section{Budget}
A thriving CubeSat astrophysics program is expected to cost less than many individual missions.
A CubeSat on the scale of a planetary science SIMPLEX CubeSat mission costs of order \$5.6M in 2016 \cite{spann_nasa_2016}.
Nearly a dozen such missions could fit within a \$65M Small Explorer budget, broadening space hardware development and funding to more PIs  across new institutions.

Margin management for CubeSat programs is import to maintaining a healthy success rate. Since CubeSats have largely been led by junior researchers and student, the full cost of engineering is often underestimated, which can lead to schedule slips and mask the true cost of the spacecraft. 
CubeSats play a critical role in educating graduate and undergraduate students.
However, it is important to ensure continuity and retention of expertise, and we encourage the exploration of a system to fund core CubeSat staff at universities and NASA centers, creating stable teams between missions.

\section*{Acknowledgements}
We would like to thank Ashley Carlton for significant contributions to early formulation of this work.

\bibliographystyle{apalike}
\bibliography{sample}

\end{document}